\documentclass[preprint,preprintnumbers,amsmath,amssymb]{revtex4}
\usepackage{graphicx}
\usepackage{chngcntr}
\usepackage{bm}
\usepackage{epsfig}
\begin{document}

\title  {Magnetic Dipole Transitions of $B_c$ and $B_c^*$ mesons in the Relativistic Independent Quark Model}
\author {Sonali Patnaik$^{1}$, P. C. Dash$^{1}$, Susmita Kar$^{2}$\footnote{email address:skar09.sk@gmail.com}, 
Sweta P. Patra$^{2}$, N. Barik$^{3}$}

\affiliation{$^{1}$ Department of Physics, Siksha O Anusandhan University, 
Bhubaneswar-751030,\\
$^{2}$ Department of Physics, North Orissa University, Baripada-757003,\\ 
$^{3}$ Department of Physics, Utkal University, Bhubaneswar-751004}

\begin{abstract}
We study M1-transitions involving mesons: $B_c(1s)$, $B_c^*(1s)$, $B_c(2s)$, $B_c^*(2s)$, $B_c(3s)$ and $B_c^*(3s)$ in the relativistic independent quark (RIQ) model based on a flavor independent average potential in the scalar-vector harmonic form. The transition form factor for $B_c^*\to B_c\gamma$ is found to have analytical continuation from spacelike to physical timelike region. Our predicted coupling constant $g_{B_c^* B_c}$ = 0.34 Ge$V^{-1}$  and decay width $\Gamma(B_c^*\to B_c\gamma)$ = 23 eV agree with other model predictions. In view of possible observation of $B_c$ and $B_c^*$  s-wave states at LHC and Z-factory and potential use of theoretical estimate on M1-transitions, we investigate the allowed as well as hindered transitions of orbitally excited $B_c$-meson states and predict their decay widths in overall agreement with other model predictions. We consider the typical case of $B_c^*(1s)\to B_c(1s)\gamma$,  where our predicted decay width which is found quite sensitive to the mass difference between $B_c^*$ and $B_c$ mesons may help in determining the mass of $B_c^*$ experimentally.
\end{abstract}
	
\maketitle

\section{Introduction}
Since its discovery at Fermilab by CDF Collaboration \cite {A1}, $B_c$-meson has aroused a great deal of interest both theoretically and experimentally due to its characteristic special features. The mesons in the 
the bottom-charm ${\bar b}c$ $(B_c)$ family lie intermediate in mass and size 
between the ${\bar c}c$ $(J/\psi)$ and ${\bar b}b (\Upsilon)$ family where the 
heavy quark interactions are believed to be understood rather well. Unlike the hidden 
flavoured heavy charmonia $({\bar c}c)$ and bottomonia $({\bar b}b)$, $B_c$-meson 
is the only lowest bound state of two different heavy quarks with open flavors (b and c) which forbid
its annhilation to photons and gluons. The ground state $B_c$ meson can therefore decay 
weakly through ${\bar b}\to {\bar c}W^+$, $c\to sW^+$ or deacy radiatively through 
$b\to b\gamma$ and ${\bar c}\to {\bar c}\gamma$ at the quark level. These decays are 
free from uncertainites which are expected in the strong decay of $B_c$-mesons and 
therefore weak and radiative decays are theoretically more tractable. The lifetime of ground state $B_c$-mesons has been carefully studied in \cite {A2,A3,A4,A5,A6}. The excited $B_c$- states lying between B-D threshold can also undergo radiative and hadronic transitions to their lower excited and ground states yielding to a rich spectroscopy of the 
radial and orbital excitations, which are more stable than their charmonium and bottomonium 
analogues. $B_c$-meson states thus provides a unique window into heavy quark dynamics and scope for independent test of quantum chromo-dynamics.

The experimental data on $B_c$-meson family are scant and data for ground state $B_c^*$ meson have not yet been possible. As estimated in \cite {A7,A8,A9} the ground state $B_c$ meson has been observed at the hadron collider, TEVATRON \cite {A10,A11} and its lifetime has been experimentally 
measured \cite {A12,A13,A14,A15} using decay channels: $B_c^{\pm}\to J/\psi l{^\pm}{\bar \nu}_e$, and 
$B_c^{\pm}\to J/\psi\pi^{\pm}$. LHCb collaboration have observed a more precise lifetime 
for $B_c^{\pm}$ mesons \cite {A16} using the decay mode $B_c\to J/\psi\mu\nu_{\mu}X$, where X denotes 
any possible additional particle in the final state. Recently the ATLAS collaboration at LHC have also detected the excited $B_c$ meson state \cite {A17} through the decay channel :
$B_c^{\pm}(2s)\to B_c^{\pm}(1s)\pi^+ \pi^-$ by using $4.9fb^{-1}$ of 7 TeV and $19.2fb^{-1}$ 
of 8 TeV pp-collision data which gives the $B_c^{\pm}$(2s) state mass $6842\pm 4\pm 5$MeV. 
It is therefore reasonable to expect a detailed study on the $B_c$ family at LHC.
But it has not been possible due to the messy QED background of the hadron collider which
contaminates the environment and make detection and precise measurements on other members 
of $B_c$ family and even the ground state $B_c^*$-meson almost impossible. In this respect 
the proposed Z-factory, an $e^+ e^-$ collider is preferred over the hadron collider at LHC. 
This is because of sufficiently high luminosity and relatively clean background offered by the
$e^+e^-$ collider that runs at Z-boson pole. Hence Z-factory is expected to enhance the 
event-accumulation rate so that $B_c$-meson excited states and possibly $B_c^*$-meson states 
are likely to be observed in near future. A possible measurement of radially excited states
of the $B_c$ family via $B_c(ns)\to B_c\pi\pi$ at LHC and the Z-factory has been discussed \cite {A18}.
However the splitting between $B_c(1s)$ and its nearest member in the $B_c$ family i.e., $B_c^*(1s)$ 
due to possible spin-spin interaction, which has been estimated \cite {A19} in the range 
$30\le\Delta m\le 50$MeV, forbids the decay mode $B_c^*\to B_c+\pi^0(\eta\eta^{\prime})$ by
energy-momentum conservation. Therefore the dominant decay mode in this sector would be the 
magnetic dipole transition:$B_c^*\to B_c\gamma$. It is worthwhile to go for a precise measurement and analysis of M1 transitions of $B_c$ and $B_c^*$ which would yield the $B_c$-spectrum and distinguish its exotic states.

The study of exclusive hadronic decays involving the non-perturbative hadronic matrix elements 
is non-trivial. Since rigorous field theoretic formulation with a first principle application 
of QCD for reliable estimation of the hadronic matrix element has not so far been possible, 
most of the theoretical attempts take resort to phenomenological approaches to probe the 
non-perturbative QCD dynamics. Different theoretical attempts \cite {A19,A20,A21,A22,A23,A24,A25,A26,A27,A28,A29,A30,A31,A32,A33} including various versions of potential models based on Bethe-Salpeter (BS) approach, light front quark (LFQ) model and QCD sum rules etc. have been 
employed to evaluate the $B_c$-spectrum and predict the mass, lifetime and decay widths of 
the ground and excited $B_c$ and $B_c^*$ meson states. We have predicted decay widths of several M1 transitions $V\to P\gamma$ and $P\to V\gamma$ in the light and heavy flavor sector in the framework of the relativistic independent quark (RIQ) model within and beyond the static approximations \cite {A34,A35}. The predicted decay widths in the light and heavy flavor sector are found to be in good agreement with other model predictions and experimental data. 
In our recent analysis \cite {A36} we studied the $q^2$ dependence of spacelike and timelike transition form-factors for energetically possible M1-transitions of heavy flavored mesons 
$(D^*,D_s^*, J/\psi)$ and $(B^*, B_c^*,\Upsilon)$ and our predicted decay widths are found compatible with the observed data and other model predictions. 
Similar studies on M1 transitions of mesons in the  $B_c$ family has not yet been undertaken in this model. Further more, with the possibility of large statistics of $B_c$ meson events at LHCb and Z-factory in near future, it is worthwhile to undertake such studies involving $B_c$- and $B_c^*$-meson ground and excited states. 

In principle one could discuss decay modes involving higher excited and P- and D- wave states of the $B_c$ family. But because their production rates are much lower and experimental measurements would be much more difficult, we do not intend to include such decay modes in this work. On the other hand we would like to analyze various possible radiative decays of the ground and 
radially excited meson states in the $B_c$ family such as $B_c^*(ns)\to B_c(ns)\gamma$;   
$B_c^*(2s)\to B_c(1s)\gamma$; $B_c^*(3s)\to B_c(2s)\gamma$; $B_c^*(3s)\to B_c(1s)\gamma$;
$B_c(2s)\to B_c^*(1s)\gamma$; $B_c(3s)\to B_c^*(2s)\gamma$ and $B_c(3s)\to B_c^*(1s)\gamma$. 
The applicability of this model has already been tested in describing a wide ranging hadronic phenomena  including the radiative, weak radiative, rare radiative \cite {A34,A35,A36,A37,A38}, leptonic \cite {A39}, weak leptonic \cite {A40}, semileptonic \cite {A41}, radiative leptonic \cite {A42}, and non-leptonic \cite {A43} decays of hadrons in the light and heavy flavor sector. Our prediction on magnetic dipole transitions of $B_c$- and $B_c^*$- meson states in this work would not only be useful for future experiments in this sector but would pin down RIQ model as a successful phenomenological model of hadrons.

The paper is organized as follows: In section 2
we present a brief account of the RIQ model. Section-3 describes model expressions for the 
transition form factors and decay width $\Gamma(V\to P\gamma)$ and $\Gamma(P\to V\gamma)$.
In section-III we discuss $q^2$-dependence of the transition form factor and numerical 
results on the coupling constants and decays rates. Section V encompasses our summary 
and conclusion.

\section{Model FrameWork}

In the RIQ model a meson is pictured as a colour-singlet assembly of a quark and an antiquark 
independently confined by an effective and average flavor independent potential in the form \cite {A34,A35,A36,A37,A38,A39,A40,A41,A42,A43}:
\begin{equation} 
U(r)=\frac{1}{2}(1+\gamma^0)(ar^2+V_0),
\end{equation}
where (a, $V_0$) are potential parameters. It is believed that the zeroth order quark dynamics 
generated by the phenomenological confining potential $U(r)$ can provide adequate tree 
level description of the decay process: 
$B_c^*\to B_c\gamma$. With the interaction potenial $U(r)$ in scalar-vector harmonic form, put into the zeroth order quark lagrangian density, the ensuing Dirac eqaution admits static solution of positive and 
negative energy. The quark orbitals so obtained correspond to all possible eigen-modes
which are described in the Appendix. 

The decay process: $B_c^*\to B_c\gamma$ in fact occurs physically in the definite momentum 
eigen-states of the participating mesons. It is therefore worthwhile to construct the 
meson states in the form of suitable wave packets reflecting appropriate momentum distribution 
between quark and antiquark in the corresponding spin-flavor configuration for which the 
individual momentum probability amplitudes $G_b({\vec p}_b)$ and ${\tilde G}_c({\vec p}_c)$
for the quark and antiquark have been obtained 
in this model via momentum projection of the bound quark orbitals. The model expression for momentum probability amplitudes are also described in the Appendix. From momentum probability amplitude of the quark and antiquark an effective momentum profile function ${\cal G}_{B_c}({\vec p_b},{\vec p_c})$ 
for a quark(b) antiquark ($\bar c$) pair is considered here in the form \cite {A34,A35,A36,A37,A38,A39,A40,A41,A42,A43}::
\begin{equation}
{\cal G}_{B_c}({\vec p_b},{\vec p_{\bar c}})=\sqrt{G_b(\vec p_b){\tilde G}_{\bar c}(\vec p_{\bar c})}
\end{equation}
in a straight forward extension of the ansatz of Margolis and Mendel in their bag model analysis \cite {A44}. 
Using ${\cal G}_{B_c}({\vec p_b},{\vec p_c})$, the meson state $|B_c(\vec P)>$ at definite 
momentum $\vec P$ and spin $S_B$ in the form of a wave packet reflecting the momentum and spin distribution among the constituent quark (b) and antiquark ($\bar c$) is constructed as
\begin{equation}
|B_c(\vec P)>={\hat \Lambda_{B_c}(\vec P,S_B)}|(\vec p_b,\lambda_b);(\vec p_c,\lambda_c)>
\end{equation}
where, $|(\vec p_b,\lambda_b);(\vec p_c,\lambda_c)>= \hat b_b^\dagger (\vec p_b,\lambda_b) 
{\hat {\tilde{b}}}_c^\dagger (\vec p_c,\lambda_c)|0>$ is a Fockspace representation of the 
unbound quark(b) and antiquark $(\bar c)$ in a 
color-singlet configuration with their respective momentum and spin as  $(\vec p_b,\lambda_b)$ and $(\vec p_c,\lambda_c)$. 
Here $\hat b_b^\dagger (\vec p_b,\lambda_b)$ and  ${\hat {\tilde{b}}}_c^\dagger (\vec p_c,\lambda_c)$
are respectively the quark and antiquark creation operators. ${\hat \Lambda}_{B_c}(\vec P,S_B)$ 
represents an integral operator: 
\begin{equation}
{\hat \Lambda}_{B_c}(\vec P,S_B)=\frac{\sqrt 3}{\sqrt{N_{B_c}(\vec P)}}\;
\sum _{{\delta_b},{\delta_{\bar c}}}\zeta_{b,{\bar c}}^{B_c}(\lambda_b,\lambda_{\bar c})
\int d^3{\vec p}_{b}\;d^3{\vec p}_{\bar c}\;\delta^{(3)}(\vec p_{b}+\vec p_{\bar c}-\vec p)
{\cal G}_{B_c}(\vec p_{b},\vec p_{\bar c})
\end{equation}
Here $\sqrt 3$ is the effective color factor, 
$\zeta_{b,{\bar c}}^{B_c}(\lambda_b,\lambda_{\bar c})$
stands for SU(6)-spin flavor coefficients for the meson $B_c(b\bar c)$.
$N(\vec P)$ is the meson-state normalization which is realized from
$<{B_c}(\vec P)\mid {B_c}({\vec P}\;^{\prime})>=\delta ^{(3)}(\vec p-{\vec p}\;^{\prime})$ 
in an integral form
\begin{equation}
N(\vec P)=\int d^3{\vec p}_b\;\mid {\cal G}_{B_c}({\vec p}_b,\vec p-{\vec p}_b)\mid ^{2}
\end{equation}
In the meson state $|{B_c}({\vec P})>$ represented by momentum wave packets of the bound quark-antiquark pair, the bound state character is thought to be embedded in the  
momentum profile function $ {\cal G}_{B_c}({\vec p}_b,\vec P-{\vec p}_b)$ used in the 
integral operator ${\hat \Lambda}_{B_c}(\vec P,S_B)$. Any residual internal dynamics 
responsible for ultimate decay process can then be studied at the level of otherwise 
free quark (b) and antiquark$(\bar c)$ using the  Feynman diagrams. 
The total contributions from appropriate Feynman diagrams is finally operated upon by 
a bag like integral operator ${\hat \Lambda}_{B_c}(\vec P,S_B)$ so as to obtain the effective 
transition amplitude for $B_c^*\to B_c\gamma$ as
\begin{equation}
S_{fi}^{B_c} = {\hat \Lambda}_{B_c}(\vec P,S_B) S_{fi}^{b\bar c}
\end{equation}
Here $S_{fi}^ {b\bar c}$ is the S-matrix elements at the constituent level 
describing $(b\bar c)\to (b\bar c) +\gamma$ and $S_{fi}^{B_c}$ is the effective 
meson-level S-matrix element describing $B_c^*\to B_c\gamma$

\section{Transition Amplitude, Transition Form Factor and Decay Width}
The hadronic matrix element for M1 transition:$B_c^*\to B_c\gamma$ can be expressed
in terms of transition form factor $F_{B_c^*B_c}(q^2)$ through the covariant expansion:
\begin{equation}
<B_c(P^{\prime})|J^{\mu}_{em}|B_c^*(P,h)>=ie\epsilon^{\mu\nu\rho\sigma}\epsilon_{\nu}
(P,h)(P+P^\prime)_{\rho}(P-P^\prime)_{\sigma}F_{B_cB_c^*}(q^2)
\end{equation}
where, $q=(P-P^{\prime})$ is the four momentum transfer, $\epsilon_{\nu}(P,h)$
is the polarization vector of vector meson $B_c^*$ with four momentum {\it P}
and helicity {\it h} and $P^{\prime}$ is the four momentum of pseudoscalar meson $B_c$. The timelike part of the covariant expansion infact vanishes 
in the $B_c^*$-meson rest frame. Hence the transition form factor 
$F_{B_c^*B_c}(q^2)$ can be calculated in RIQ-model from the non-vanishing spacelike 
part of hadronic matrix element (7) using the appropriate meson states as in (3-4). 
In the $B_c^*$-meson rest frame:
$q^2=M^2_{B_c^*}+M^2_{B_c}-2M_{B_c^*}\sqrt{{\vec k}^2+ M^2_{B_c}}$ has a kinematic range: $0\le(q^2)\le(M_{B_c^*}-M_{B_c})^2$, where ${\vec k}$ is third momentum of emitted photon.
\begin{figure}
	\begin{center}
		\includegraphics[width=12 cm,height=4 cm]{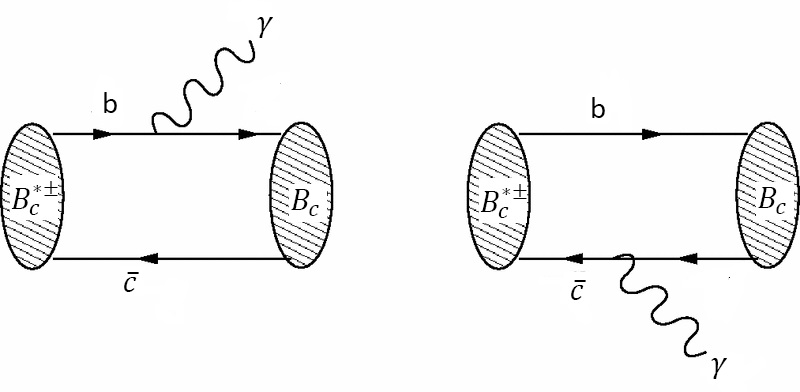}
	\end{center}
	\caption{{Lowest order Feynman diagram contributing to $B_c^*$ radiative transition}}
\end{figure}
Now assuming the decay process: 
$B_c^*\to B_c\gamma$, depicted in the lowest order Feynman diagrams [Fig. 1(a,b)], is predominantly a single vertex decay process governed mainly by photon emission from independentely confined quark or antiquark inside the meson, the S-matrix element 
in the configuration space can be written as 
\begin{equation}
S_{B_cB_c^*}=<B_c\gamma |-ie\int d^4xT[\sum_qe_q{\bar \psi}_q(x)\gamma^\mu\psi_q(x)
A_\mu (x)]|B_c^*>
\end{equation}
which can be reduced to
\begin{equation}
S_{B_cB_c^*}=i\sqrt{\alpha/E_k}<B_c(P^{\prime})|\sum_{q,\lambda,\lambda^{\prime}}
\frac{e_q}{e}\int\frac{dp\;dp^{\prime}}{\sqrt{4E_pE_{p^\prime}}}\delta^{(4)}
(p^{\prime}+k-p){\cal D}(p^\prime\lambda^{\prime};p\lambda;k\delta)|B_c^*(P)>
\end{equation}
where,
\begin{eqnarray}
{\cal D}(p^\prime\lambda^{\prime};p\lambda;k\delta)=
{\bar U}(p^\prime,\lambda^{\prime})\gamma.\epsilon(k,\delta)U(p,\lambda)
b_q^{\dagger}(p^\prime,\lambda^{\prime})b_q(p,\lambda)\nonumber\\
-{\bar V}(p,\lambda)\gamma.\epsilon(k,\delta)V(p^\prime,\lambda^{\prime})
{\tilde b}_q^{\dagger}(p^\prime,\lambda^{\prime}){\tilde b}_q(p,\lambda)
\end{eqnarray}
Here $\alpha$ is fine structure constant, k and $E_k$  are four momentum and energy of the emitted photon;   $E_P=M_{B_c^*}$ and $E_{P^{\prime}}$ are energies  
of intial and daughter meson, respectively.
Then using appropriate wavepackets representing the meson states 
($|B_c^*>$, $|B_c>$) and explicit forms of Dirac spinors: ${\hat U}(p_b,\lambda_b)$
and ${\hat V}(p_c,\lambda_c)$, the S-matrix element in $B_c^*$-meson rest frame is obtained as
\begin{equation}
S_{B_cB_c^*}=i\sqrt{\alpha/k_0}\delta^{(4)}(P^{\prime}+k-\hat{O}M_{B_c^*})
[Q(P^{\prime},{\vec k})-\tilde{Q}(P^{\prime},{\vec k})]
\end{equation}
Here $P^{\prime}\equiv(E_p,{\vec p}^{\prime})$; $\hat{O}\equiv(1,0,0,0)$,
${\vec P}^{\prime}+{\vec k}=0$
\begin{eqnarray}
Q({\vec k})=\sum\frac{e_{q_1}}{e}\zeta^{B_c^*}_{b,c}(\lambda_b\lambda_c)
\zeta^{B_c}_{b,c}(\lambda_b^{\prime}\lambda_c)\chi^{\dagger}_{\lambda_b^{\prime}}
({\vec \sigma.\vec K})\chi_{\lambda_b}
J_b(\vec k)\nonumber\\
{\tilde Q}({\vec k})=\sum\frac{e_{q_2}}{e}\zeta^{B_c^*}_{b,c}(\lambda_b\lambda_c)
\zeta^{B_c}_{b,c}(\lambda_b\lambda_c^{\prime})
{\tilde \chi}^{\dagger}_{\lambda_c}({\vec \sigma.\vec K})
{\tilde \chi}_{\lambda_c}^{\prime}
J_c(\vec k)
\end{eqnarray}
with ${\vec K}={\vec k}\times {\vec \epsilon}({\vec k},\delta)$ and 
\begin{eqnarray}
J_b=\int d{\vec p_b}\frac{{\cal G}_{B_c^*}({\vec p_b},
	-{\vec p_b}){\cal G}_{B_c}({\vec p_b}-{\vec k}, -{\vec p_b})}
{\sqrt{{{\bar N}_{B_c^*}(0){\bar N}_{B_c}(\vec k)}}}
\sqrt{{(E_{p_b}+m_b)\over
		{4E_{p_b}}E_{p_bk}(E_{p_bk}+m_b)}}\nonumber\\
J_c=\int d{\vec p_c}\frac {{\cal G}_{B_c^*}({\vec p_c},
	-{\vec p_c})\;{\cal G}_{B_c}(-{\vec p_c}, {\vec p_c}-{\vec k})}
{\sqrt{{{\bar N}_{B_c^*}(0){\bar N}_{B_c}(\vec k)}}}
\sqrt{{(E_{p_c}+m_c)\over {4E_{p_c}E_{p_ck}(E_{p_ck}+m_c)}}}
\end{eqnarray}
We denote $E_{p_{b,c}}=\sqrt{{\vec p}^2_{b,c}+m^2_{b,c}}$ and
$E_{p_{b,c}k}=\sqrt{({\vec p}^2_{b,c}-{\vec k})^2+m^2_{b,c}}$
and use the so called loose binding approximation: $E_{p_b}+E_{p_c}=M_{B_c^*}$
and $E_{p_bk}+E_{p_c}=E_{p_b}+E_{p_ck}=E_{B_c}$ here to ensure energy conservation at the photon
hadron vertex.

Then specifying appropriate spin flavor-coefficients 
$\zeta^{B_c^*}_{b,c}(\lambda_b\lambda_c)$ and $\zeta^{B_c}_{b,c}(\lambda_b\lambda_c)$
for the vector and pseudoscalar mesons, the invariant transition amplitude is 
extracted from the S-matrix elements (11) in the form:
\begin{equation}
{\cal M}_{B_cB_c^*}=\sqrt{4\pi\alpha}\sqrt{2M_{B_c^*}2E_{B_c}}\;F_{B_cB_c^*}
(\vec k)K_{S_V}
\end{equation}  
Similarly for transition $B_c\to B_c^*\gamma$, the invariant transition amplitude can also be obtained in the form of form factor $F_{B_c^*B_c}(\vec k)$. Here $K_{S_V}$ for both the decay modes corresponding to spin statess $(\pm 1,0)$ stand for 
\begin{eqnarray}
K_{S_V}(B_c^*\to B_c\gamma)=\left [\mp(K_1\pm iK_2)/\sqrt{2},K_3\right ]
\nonumber\\
K_{S_V}(B_c\to B_c^*\gamma)=\left [\pm(K_1\mp iK_2)/\sqrt{2},K_3\right ]
\end{eqnarray}
Note that a sum over photon polarization index $\delta $ and vector meson 
spin states $(\pm 1,0)$ yields a general relation
\begin{equation}
\sum_{\delta,S_V}|K_{S_V}|^2=2k^2
\end{equation}
Then the decay widths for $B_c^*\to B_c\gamma$ and $B_c\to B_c^*\gamma$ are 
obtained from the generic expression:
\begin{equation}
\Gamma =\frac{1}{(2\pi)^2}\frac{1}{2M_{B_c^*,B_c}}\int\frac{d^3P^{\prime}
	d^3k}{2E_{P^{\prime}}2E_k}{\bar \sum}|{\cal M}_{fi}|^2
\delta ^{(4)}(P^{\prime}+k-{\hat O}M_{B_c^*,B_c})
\end{equation}
in the form:
\begin{eqnarray}
\Gamma (B_c^*\to B_c\gamma)=\frac{\alpha}{3}{\bar k}^3\left |\sqrt{E_{B_c}
	(\vec k)/M_{B_c^*}}\;F_{B_cB_c^*}(q^2)\right |^2\nonumber\\ 
\Gamma (B_c\to B_c^*\gamma)=\alpha{\bar k}^3\left |\sqrt{E_{B_c^*}(\vec k)/
	M_{B_c}}\;F_{B_c^*B_c}(q^2)\right |^2
\end{eqnarray}
It may be mentioned that a phase space factor such as $\sqrt{E_{B_c}(\vec k)/M_{B_c^*}}$ is arising here out of the argument factorization of energy delta function which has been extracted from the constituent level 
integration (11) under certain approximation in order to realize correct 
photon energy at the mesonic level. Infact starting  with a relativistic effective interaction of the form $F_{VP}(q^2)\epsilon^{\mu\nu\rho\sigma}\partial_{\mu}
A_{\nu}(x)\partial_{\rho}V_{\sigma}(x)P(x)$ where $A_{\nu}(x)$, $V_{\sigma}(x)$  and  $P(x)$ 
are,respectively the fields of photon, vector meson and pseudoscalar meson, one can arrive at the expression for $\Gamma(V\to P\gamma)$ in terms of transition form factor $F_{VP}(q^2)$ without the mesonic level phase-space factor. 
The spurious phase space factor arising here is not a problem typical to this model calculation. It is indeed a pathological problem common to all phenomenological models attempting to explain the hadronic level decays in terms of constituent level dynamics considered in zeroth order. However an explicit cancelletion of such phase space factor taken approximately along with the contribution of quark spinors have been obtained by authors \cite {A45} within the scope of their models. Here we would like to push back the phase space factor from the mesonic level to quark level integral 
$J_q(\vec k)$ describing $F_{B_c^*B_c}(\vec k)$ under the same approximation with which it was extracted out through the argument factorization of 
energy delta function. The phase space factor $\sqrt{E_{B_c}({\vec k})
	/M_{B_c^*}}$ taken in the form $\sqrt{\frac{(E_{p_{b,c}k}+E_{p_{c,b}})}
	{(E_{p_b}+E_{p_c})}}$
into the quark level integral in Eq. (13); reduces $J_{b,c}(\vec k)$ to 
$I_{b,c}(\vec k)$ yielding

\begin{eqnarray}
I_b(\vec k)=\frac{1}{\sqrt{{\bar N(0)\bar N(\vec k)}}}
\int d{\vec p_{q_1}}{\cal G}_{B_c^*}({\vec p_b},-{\vec p_b})
{\cal G}_{B_c}({\vec p_b}-{\vec k},-{\vec p_b})
\sqrt{{(E_{p_b}+m_b)(E_{{p_b}k}+E_{p_c}})\over
	{4E_{p_b}}E_{{p_b}k}(E_{{p_b}k}+m_b)(E_{p_b}+E_{p_c})}
\nonumber\\
I_c(\vec k)=\frac{1}{\sqrt{{\bar N(0)\bar N(\vec k)}}}
\int d{\vec p_{q_1}}{\cal G}_{B_c^*}(-{\vec p_c},
-{\vec p_c}){\cal G}_{B_c}(-{\vec p_c}, {\vec p_c}-{\vec k})
\sqrt{{(E_{p_c}+m_c)(E_{{p_c}k}+E_{p_b}})\over 
	{4E_{p_c}E_{{p_c}k}(E_{{p_c}k}+m_c)(E_{p_b}+E_{p_c})}}
\end{eqnarray}
in terms of which the transition form factor is found to be:
\begin{equation}
F_{B_cB_c^*}(\bar k)=\frac{1}{3}[2I_c(\bar k)-I_b(\bar k)]
\end{equation}

Finally the decay widths for transitions: $B_c^*\to B_c\gamma$ and
$B_c^*\to B_c\gamma$ are obtained in the usual form:
\begin{eqnarray}
\Gamma (B_c^*\to B_c\gamma)=\frac{\alpha}{3}{\bar k}^3\left 
|g_{B_c^*B_c}(\bar k)\right |^2\nonumber\\
\Gamma (B_c\to B_c^*\gamma)=\alpha{\bar k}^3\left 
|g_{B_c^*B_c}(\bar k)\right |^2
\end{eqnarray}
where, $\bar k=\frac{(M^2_{B_c^*}-M^2_{B_c})}{2M_{B_c^*}}$ is the energy of outgoing photon; $g_{B_cB_c^*}(\bar k)$
and $g_{B_c^*B_c}(\bar k)$ are coupling constants obtained from respective
transition form factor in the limit $q^2\to 0$ that corresponds to real photon.
We consider here the transverse $(h=\pm 1)$ polarization only to get the
coupling constant since the longitudinal component of vector meson does not convert into a real photon.

\section{Numerical results and discussion}
For numerical analysis of radiative decay of the ground state $B_c^*(1s)$
meson, we take the quark masses $m_q$, corresponding binding energies $E_q$ and potential parameters (a,$V_0$) as those fixed from hadron spectroscopy by fitting the data of heavy quarkonia \cite {A46} and then used to describe a wide ranging hadronic phenomena \cite {A34,A35,A36,A37,A38,A39,A40,A41,A42,A43} as
\begin{eqnarray}
(a, V_0)&\equiv & (0.017166\;{GeV}^3, -0.1375\;GeV)\nonumber\\
(m_b, m_c, E_b, E_c)&\equiv & (4.77659, 1.49276, 4.76633, 1.57951)\;GeV
\end{eqnarray}
\begin{figure}
	\begin{center}
		\includegraphics[width=14 cm,height=10 cm]{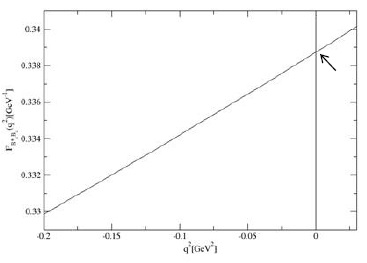}
	\end{center}
	\caption{{Dependence of $\Gamma(B_c^*\to B_c\gamma)$ on $\Delta m = M_{B_c^*}-M_{B_c}$}}
\end{figure}
Since the mass of $B_c^*(1s)$-meson has not yet been observed, we take our predicted values; $M_{B_c}=6.2642$ GeV and $M_{B_c^*}=6.3078$ GeV \cite {A40}. Note that our predicted value of $M_{B_c}$ is close to the central value of observed one i.e., $M_{B_c}^{expt}=6.2751$ GeV \cite {A47}. 
In Fig.2 we depict the $q^2$- dependence of form factor $F_{B_c^*B_c}(q^2)$ and show its analytical continuation from the spacelike $(q^2<0)$ region to the physical timelike $(0\le q^2\le q^2_{max})$ region.
Here $q^2_{max}=(M_{B_c^*}-M_{B_c})^2$ 
corresponds to zero recoil point for the $B_c$ meson which is shown by the arrow in Fig. 2. The coupling constant $g_{B_c^*B_c}$ for real photon case is 
calculated from the expression of the form factor $F_{B_c^*B_c}(q^2)$ in the limit
$q^2\to 0$ where the final state $B_c$ meson gets recoiled with maximum three momentum $|\bar k|=\frac{(M^2_{B_c^*}-M^2_{B_c})}{2M_{B_c^*}}$. Our prediction
$g_{B_c^*B_c}=0.34$ $GeV^{-1}$ is comparable to the results of $0.273 [0.257] GeV^{-1}$
for linear [HO] potential from LFQM \cite {A32} and $0.27\pm0.095$ $GeV^{-1}$ from QCD sum rule 
approach \cite {A33}.

\begin{table}[ht]
	\renewcommand{\arraystretch}{1.3}
	\setlength\tabcolsep{5pt}
	\begin{center}
		\caption{{\bf Predicted transition energy, coupling constant and decay width in the RIQ model.}}
		\begin{tabular}{|c|c|c|c|}
			\hline
			\hline
			Transitions & Transition Energy (MeV)& Coupling Constant $(GeV^{-1})$ & Decay Width (KeV)\\
			\hline
			$1^{3}S_1\to 1^{1}S_0$ & 0.04344 &0.3392   & 0.023\\
			$2^{3}S_1\to 2^{1}S_0$ & 0.06806 &0.300066 & 0.069\\
			$3^{3}S_1\to 3^{1}S_0$ & 0.12285 &0.338143 & 0.516\\
			$2^{2}S_1\to 1^{1}S_0$ &0.61589  &0.02609   & 0.387\\
			$3^{3}S_1\to 2^{1}S_0$ &0.40559  &0.02919   & 0.138\\
			$3^{3}S_1\to 1^{1}S_0$ &0.927079 &0.01121   & 0.244\\
			$2^{1}S_0\to 1^{3}S_1$ &0.51325  &0.038745  & 1.481\\
			$3^{1}S_0\to 2^{3}S_1$ &0.22174  &0.05898   & 0.277\\
			$3^{1}S_0\to 1^{3}S_1$ &0.77978  &0.04138   & 5.927\\
			\hline
		\end{tabular}
	\end{center}
\end{table} 

Finally our predicted decay width
$\Gamma(B_c^*\to B_c\gamma)=23$ eV is compatible with
other theoretical predictions such as 17 eV from Bethe-Salpeter approach \cite {A19}, 33 eV from the relativistic quark model \cite {A24}, 
59 eV from the Richardson's potential \cite {A23}, 60 eV from the non relativistic potential \cite {A21}, 80 eV from the relativized quark model \cite {A25} and $ 133.9\pm 79.7$ eV from QCD sum rule approach \cite {A33}.
\begin{figure}
	\begin{center}
		\includegraphics[width=16 cm,height=14 cm]{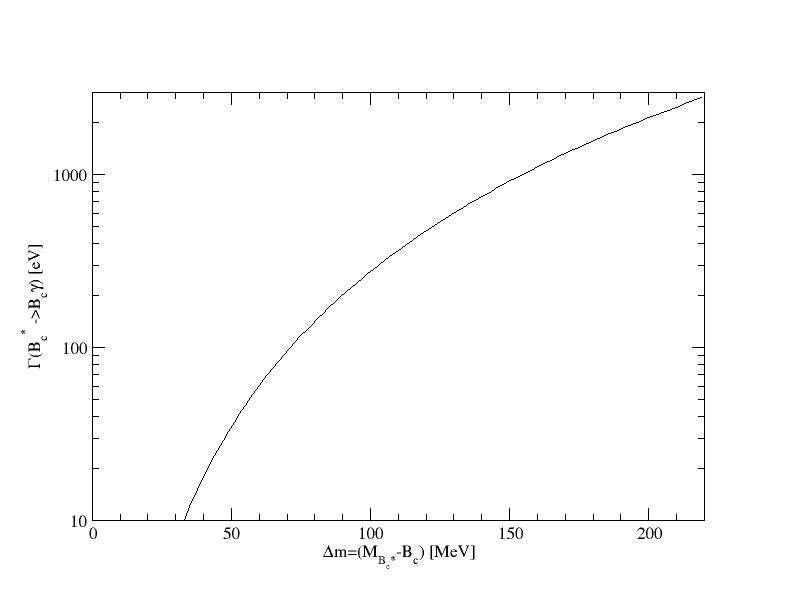}
	\end{center}
	\caption{{Dependence of $\Gamma(B_c^*\to B_c\gamma)$ on $\Delta m = M_{B_c^*}-M_{B_c}$}}
\end{figure}
\newpage
For unmeasured $B_c^*$ meson mass, we take a range of the $B_c^*$ meson mass as $33\;MeV\le\Delta m = (M_{B_c^*}-M_{B_c})\le 220\;MeV$. The lower value of $\Delta m$ chosen here corresponds to our predicted $B_c^*$ meson mass (i.e., $M_{B_c^*}=6308$ MeV). The decay width $\Gamma(B_c^*\to B_c\gamma)$ being proportional to $\Delta m^3=(M_{B_c^*}-M_{B_c})^3$ is found quite sensitive to $B_c^*$ meson mass as depicted in Fig.3. In the same range of $\Delta m$ our predicted decay width is found to vary widely from 0.23 eV to 2824.28 eV. This is comparable to predicted values in the range:
22.4[19.9] eV $\sim 1836[1631]$ eV for $\Delta m = 50 MeV$ $\sim 220$ MeV obtained for linear [HO] potential in LFQ model \cite {A32}. The sensitivity of $\Gamma(B_c^*\to B_c\gamma)$ on $B_c^*$ mass in this model provide a clue for experimental determination of $B_c^*$-mass which is expected at LHCb and Z-factory in near future. 

For numerical analysis of transitions involving radially excited $B_c$ and $B_c^*$ mesons, we take the same quark masses and potential parameter as in (22). The quark and antiquark binding energies for radially excited states (2s and 3s) are obtained in this model by solving the corresponding cubic equations for n=2 and 3 representing their bound states conditions. The binding energies for quark b and antiquark ${\bar c}$ are found to be:
\begin{eqnarray} 
( E_b ; E_c ) = ( 5.05366 ; 1.97016 ) GeV \nonumber\\
( E_b ; E_c ) = ( 5.21703 ; 2.22479 ) GeV 
\end{eqnarray}
for 2s and 3s states, respectively.
With the model parameters (a,$V_0$) and quark mesons $m_q$ as in (22) and binding energies $E_q$ obtained in the model as shown in (23), we generate the mass splitting as done in \cite {A46} between 
$B_c^*$ and $B_c$ mesons in 2s-states yielding $M_{B_c}^* = 6.88501$ GeV and 
$M_{B_c}= 6.78521$GeV. Our predicted mass $M_{B_c}(2s)$ for example is found  57 MeV below the observed value of $6842\pm 4\pm 5$ MeV \cite {A17}. We thus encounter a difficulty here to make sure all the meson states (ground and excited) to have their respective correct masses with same set of input parameters. This is indeed a problem common to all potential models especially for states above the threshold. Just as in all other potential models, we too cannot expect to obtain precise meson masses for all the states. So we adjust the $V_0$ value in our potential to a new value i.e -0.01545 GeV so as to set the $B_c$(2s) mass equal to the observed value as done by T.Wang {\it{et al.}} in their analysis based on the instantaneous approximated Bethe-Salpeter approach \cite {A27}. With the newly adjusted value of $V_0$ and other relevant input parameters (22,23), we predict the mass of meson states:
$B_c^*(2s)$, $B_c(2s)$, $B_c^*(3s)$ and $B_c(3s)$ as:
\begin{eqnarray}
( M_{B_c^*}(2s); M_{B_c}(2s) ) = ( 6910.3 ; 6841.9 ) MeV\nonumber\\
( M_{B_c^*}(3s) ; M_{B_c}(3s) ) = (7259.5; 7135.6) MeV
\end{eqnarray} 
Using appropriate wave packets for initial and daughter meson states, we calculate the invariant transition matrix element from (9) and extract the coupling constants $g_{B_c^* B_c}=F_{B_c^* B_c}(q^2=0)$. Then substituting the value of 
$g_{B_c^* B_c}$ in (21), we evaluate decay widths. Our predicted coupling constants and decay widths for decay modes involving ground and radially excited states along with the associated photon energy are listed in Table-1. It can be noted here that the transition energy involved in different decay modes may differ by a factor of $2\sim 3$ but the corresponding coupling constants are found to vary only marginally. Most of our predictions on decay widths are also found in qualitative agreement with other model predictions as shown in Table-2. For M1 transition:
$B_c({2s})\to B_c^*({1s})\gamma$, although our result is found large compared to most other model predictions but it finds an order of magnitude agreement with the result of the recent work of Devlani {\it et al.} \cite {A30}. However for transitions:
$B_c^*(3s)\to B_c(3s)\gamma$ and $B_c(3s)\to B_c^*(2s)\gamma$ there is order of magnitude mismatch between our result and most other model predictions. It may be mentioned here that the mass of orbitally excited $B_c(3s)$, $B_c^*(3s)$, and $B_c^*(2s)$ states have not yet been measured. Different models use different meson masses to evaluate decay widths. Being sensitive to the value of meson masses it is not therefore surprising to have predicted decay widths varying from one model to other.

\begin{table}[ht]
	\renewcommand{\arraystretch}{1.3}
	\setlength\tabcolsep{5pt}
	\begin{center}
		\caption{{\ Comparison of theoretical predictions on M1 transition rate (KeV)}}
		\begin{tabular}{|c|c|c|c|c|c|c|c||c|}
			\hline
			\hline
			Transitions & Present Work & \cite {A25} & \cite {A19} & \cite {A24} & \cite {A21} & \cite {A20} & \cite {A23} & \cite {A30} \\
			\hline 
			$1^{3}S_1\to 1^{1}S_0$ & 0.023 & 0.08 & 0.017 & 0.033 & 0.06 & 0.135 & 0.059 & - \\
			$2^{3}S_1\to 2^{1}S_0$ & 0.069 & 0.01 & - & 0.017 & 0.01 & 0.029 & 0.012 & - \\ 
			$3^{3}S_1\to 3^{1}S_0$ & 0.516 & 0.003 & - & - & - & - & - & - \\
			$2^{2}S_1\to 1^{1}S_0$ & 0.387 & 0.6 & 0.28 & 0.428 & 0.098 & 0.123 & 0.122 & - \\
			$3^{3}S_1\to 2^{1}S_0$ & 0.138 & 0.2 & - & - & - & - & - & - \\
			$3^{3}S_1\to 1^{1}S_0$ & 0.244 & 0.6 & 0.37 & - & - & - & - & - \\
			$2^{1}S_0\to 1^{3}S_1$ & 1.481 &  0.3 & 0.38 & 0.488 & 0.096 & 0.093 & 0.139 & 1 \\
			$3^{1}S_0\to 2^{3}S_1$ & 0.277 & 0.06 & 0.25 & - & - & - & - & - \\ 
			$3^{1}S_0\to 1^{3}S_1$ & 5.927 & 4.2 & 0.074 & - & - & - & - & - \\ 
			\hline
		\end{tabular}
	\end{center}
\end{table} 

\newpage  

The transitions of the type $B_c^*(ns)\to B_c(ns)\gamma$ are known as allowed transitions where as the transitions in which principal quantum numbers change, are referred to as hindered ones. In theoretical studies \cite {A20,A21,A23} based on non-relativistic approach, the M1-transitions especially hindered ones have been predicted to have large decay widths. Introducing relativistic effect into the analysis \cite {A24} the results are found to be rather small. Infact the relativistic corrections are implicitly taken into account by invoking spin-spin interactions while extracting the wave functions in this model and reproducing hyperfine splitting between vector meson and its pseudoscalar counterpart. In the present study the relativistic effect on ${\bar c}$ quark which is not so heavy compared to b-quark is found to be significant. This along with our choice of interaction potential U(r) in equally mixed scalar-vector harmonic form yields the results as shown in Table 2 in qualitative agreement with other model predictions. 

\section {Summary and Conclusion}
In this work we study M1 transitions of the ground and excited s-wave states of $B_c$- and $B_c^*$- meson  in the framework of relativistic independent quark (RIQ) model based on an equally mixed scalar-vector harmonic form. We predict the $q^2$-dependence of transition form factor $F_{B_c^*B_c}(q^2)$ for the transition: $B_c^*(1s)\to B_c(1s)\gamma$, where the spacelike $(q^2< 0)$ form factor is shown to have analytical continuation to the physical timelike $(0\le\ q^2\le q_{max})$  region, with 
$q^2_{max} = (M_{B_c^*}-M_{B_c})^2$ corresponding to the zero-recoil point for the daughter meson $(B_c)$. We extract the coupling constant $g_{B_c^*B_c}$ from $F_{B_c^*B_c}(q^2)$ in the limit $q^2\to 0$ for real photon case. Our prediction for coupling constant $g_{B_c^* B_c}$ = 0.34 Ge$V^{-1}$ is comparable to the result of  0.273 [0.257] Ge$V^{-1}$  for linear [HO]  potential from LFQ model \cite {A32} and 
$0.27\pm 0.095  GeV^{-1}$ from the QCD sum rule approach \cite {A33}. We also predict decay width: $\Gamma({B_c^*(1s)\to B_c(1s)}\gamma) = 23$ eV in comparison with other theoretical predictions such as 17 eV from Bethe-Salpeter approach \cite {A19}, 33 eV from relativistic potential \cite {A24}, 60 eV from non-relativistic potential \cite {A21}, 59 eV from the Richardson's potential \cite {A23}, 80 eV from the relativized quark model \cite {A25} and $133.9\pm 79.7$ eV from the QCD sum rule approach \cite {A33}. Since the decay width: $\Gamma(B_c^*\to B_c\gamma)$ is proportional to $(\Delta m)^3$, we study the dependence of decay width on $\Delta m = M_{B_c^*} - M_{B_c}$ for which we take a range of ${\Delta m}$ values: $33 MeV\le\Delta m\le 220$ MeV. 
The lowest values of 33  MeV corresponds to our predicted $B_c^*$ mass of 6308 MeV. We find that although the value of the transition form factor $F_{B_c^*B_c}(q^2)$ is not sensitive to $B_c^*$- meson mass, the decay width $\Gamma(B_c^*\to B_c\gamma)$ is found quite sensitive to $M_{B_c^*}$ .This is quite evident from our predicted values varying widely in the range: 
$(0.23 \sim 2824.28)$ eV for $\Delta m = 33MeV \sim 220 MeV$. The sensitivity of $\Gamma(B_c^*\to B_c\gamma)$  on $B_c^*$ - meson mass would guide the experiment for measurement  of $B_c^*$ - meson mass which is expected at LHC and the proposed Z-factory in near future. 

For analysis of M1 transitions involving radially excited 2s- and 3s- wave states, we first find the binding energies of quark b and antiquark $\bar c$ by solving the cubic equation representing respective bound state condition in this model. Then by suitably adjusting the value of $V_0$ of our potential U(r) to a  new value $\sim -0.01545$ GeV,
we generate the mass splitting so as to obtain the mass of $B_c(2s)$- meson equal to its observed value \cite {A17}. The corresponding meson masses obtained in this model are: $M_{B_c^*}(2s) = 6910.3$ GeV, $M_{B_c}(2s) = 6841.9$ MeV, $M_{B_c^*}(3s) = 7259.5$ MeV and $M_{B_c}(3s) = 7135.6$ MeV. 

Finally we predict transition energies, coupling constants and decay widths for energetically possible decay modes involving $B_c^*(ns)$ and $B_c(ns)$ states with n=1,2,3. We find that the transition energy may change by a factor of about $2\sim 3$ from one transition mode to other but the corresponding coupling constant changes only marginally. Our predicted decay widths for transition involving the ground and excited $B_c$- meson s-wave states, are found mostly in qualitative agreement with other model predictions except in few cases that involve excited $B_c^*(2s)$ and $B_c^*(3s)$ and $B_c(3s)$ states. It may be mentioned here that in evaluating decay widths for transitions:
$B_c^*(3s)\to B_c(3s)\gamma$, 
$B_c(3s)\to B_c^*(2s)\gamma$ ,for example, different models use different meson masses obtained in their respective model calculations since masses of these excited states have not yet been measured. The predicted decay widths for these transitions are found to vary from one model to other as expected. The present model, within its working approximation, thus provides a realistic framework to describe M1-transitions of $B_c$ and $B_c^*$  s-wave states based on the conventional picture of photon emission induced by the quark electromagnetic current. Besides S-wave states there are two P-wave multiplets and one D-wave multiplet for the members of $B_c$- family lying below the B-D threshold, which we have not considered in this work. We would like to address this issue in our future communication.

\newpage

\section{Appendix: Quark orbitals and momentum probability amplitudes of constituent quarks}
The interaction potential $U(r)=\frac{1}{2}(1+\gamma^0)(ar^2+V_0)$ in the scalar-vector harmonic form in the RIQ model, put into the quark lagrangian density, the ensuing Dirac 
equation admits static solutions of positive and negative energies in zeroth order as
\begin{eqnarray}
\psi^{(+)}_{\xi}(\vec r)\;&=&\;\left(
\begin{array}{c}
\frac{ig_{\xi}(r)}{r} \\
\frac{{\vec \sigma}.{\hat r}f_{\xi}(r)}{r}
\end{array}\;\right)U_{\xi}(\hat r)
\nonumber\\
\psi^{(-)}_{\xi}(\vec r)\;&=&\;\left(
\begin{array}{c}
\frac{i({\vec \sigma}.{\hat r})f_{\xi}(r)}{r}\\
\frac{g_{\xi}(r)}{r}
\end{array}\;\right){\tilde U}_{\xi}(\hat r)
\end{eqnarray}
where, $\xi=(nlj)$ represents a set of Dirac quantum numbers spececifying 
the eigen-modes;
$U_{\xi}(\hat r)$ and ${\tilde U}_{\xi}(\hat r)$
are the spin angular parts given by,
\begin{eqnarray}
U_{ljm}(\hat r) &=&\sum_{m_l,m_s}<lm_l\;{1\over{2}}m_s|
jm>Y_l^{m_l}(\hat r)\chi^{m_s}_{\frac{1}{2}}\nonumber\\
{\tilde U}_{ljm}(\hat r)&=&(-1)^{j+m-l}U_{lj-m}(\hat r)
\end{eqnarray}
With the quark binding energy $E_q$ and quark mass $m_q$
written in the form $E_q^{\prime}=(E_q-V_0/2)$,
$m_q^{\prime}=(m_q+V_0/2)$ and $\omega_q=E_q^{\prime}+m_q^{\prime}$, one 
can obtain solutions to the resulting radial equation for 
$g_{\xi}(r)$ and $f_{\xi}(r)$in the form:
\begin{eqnarray}
g_{nl}&=& N_{nl} (\frac{r}{r_{nl}})^{l+l}\exp (-r^2/2r^2_{nl})
L_{n-1}^{l+1/2}(r^2/r^2_{nl})\nonumber\\
f_{nl}&=& N_{nl} (\frac{r}{r_{nl}})^{l}\exp (-r^2/2r^2_{nl})\nonumber\\
&\times &\left[(n+l-\frac{1}{2})L_{n-1}^{l-1/2}(r^2/r^2_{nl})
+nL_n^{l-1/2}(r^2/r^2_{nl})\right ]
\end{eqnarray}
where, $r_{nl}= a\omega_{q}^{-1/4}$ is a state independent length parameter, $N_{nl}$
is an overall normalisation constant given by
\begin{equation}
N^2_{nl}=\frac{4\Gamma(n)}{\Gamma(n+l+1/2)}\frac{(\omega_{nl}/r_{nl})}
{(3E_q^{\prime}+m_q^{\prime})}
\end{equation}
and
$L_{n-1}^{l+1/2}(r^2/r_{nl}^2)$ etc. are associated Laguerre polynomials. The radial solutions yields an independent quark bound-state condition in the form of a cubic equation:
\begin{equation}
\sqrt{(\omega_q/a)} (E_q^{\prime}-m_q^{\prime})=(4n+2l-1)
\end{equation}
The solution of the cubic equation provides the zeroth order binding energies of 
the confined quark and antiquark for all possible eigen modes.

In the relativistic independent particle picture of this model, the constituent quark 
and antiquark are thought to move independently inside the $B_c$-meson bound state 
with momentum $\vec p_b$ and $\vec p_c$, respectively. Their individual momentum probability 
amplitudes are obtained in this model via momentum projection of respective quark orbitals in following forms:
For ground state mesons:(n=1,l=0)
\begin{eqnarray}
G_b(\vec p_b)&=&{{i\pi {\cal N}_b}\over {2\alpha _b\omega _b}}
\sqrt {{(E_{p_b}+m_b)}\over {E_{p_b}}}(E_{p_b}+E_b)\exp {(-{
		{\vec p}^2\over {4\alpha_b}})}\nonumber\\
{\tilde G}_c(\vec p_c)&=&-{{i\pi {\cal N}_c}\over {2\alpha _c\omega _c}}
\sqrt {{(E_{p_c}+m_c)}\over {E_{p_c}}}(E_{p_c}+E_c)\exp {(-{
		{\vec p}^2\over {4\alpha_c}})}
\end{eqnarray}
For excited meson state:(n=2, l=0)
\begin{eqnarray}
G_b(\vec p_b) ={{i\pi {\cal N}_b}\over {2\alpha _b\omega _b}}
\sqrt {{(E_{p_b}+m_b)}\over {E_{p_b}}}\exp {(-{
		{\vec p}^2\over {4\alpha_b}})}\sqrt{(A^2_b+B^2_b)}e^{i\phi_b}\nonumber\\
{\tilde G}_c(\vec p_c)=-{{i\pi {\cal N}_c}\over {2\alpha _c\omega _c}}
\sqrt {{(E_{p_c}+m_c)}\over {E_{p_c}}}\exp {(-{
		{\vec p}^2\over {4\alpha_c}})}\sqrt{(A^2_c+B^2_c)}e^{i{ \phi_c}}
\end{eqnarray}
where,
\begin{eqnarray}
A_{b,c}&=&\frac{3}{\sqrt{\pi}}(E_{p_{b,c}}-m_{b,c})\sqrt{\frac{\alpha_{b,c}}{p^2_{b,c}}}
\;(3-\frac{p^2_{b,c}}{\alpha_{b,c}})\nonumber\\
B_{b,c}&=&\frac{\omega_{b,c}}{2}(\frac{p^2_{b,c}}{\alpha_{b,c}}-3)
+(E_{p_{b,c}}-m_{b,c})(1+\frac{\alpha_{b,c}}{p^2_{b,c}})
\end{eqnarray}

For the excited meson state (n=3, l=0)
\begin{eqnarray}
G_b(\vec p_b) ={{i\pi {\cal N}_b}\over {4\alpha _b\omega _b}}
\sqrt {{(E_{p_b}+m_b)}\over {E_{p_b}}}
\exp {(-{{\vec p}^2\over {4\alpha_b}})}\sqrt{(A^2_b+B^2_b)}e^{i\phi_b}\nonumber\\
{\tilde G}_c(\vec p_c) =-{{i\pi {\cal N}_c}\over {4\alpha _c\omega _c}}\sqrt {{(E_{p_c}+m_c)}\over {E_{p_c}}}\exp {(-{{\vec p}^2\over {4\alpha_c}})}\sqrt{(A^2_c+B^2_c)}e^{i{\phi_c}}
\end{eqnarray}

where,
\begin{eqnarray}
A_{b,c}&=&\frac{\omega_{b,c}}{2p_{b,c}}\sqrt{\frac{\alpha_{b,c}}{\pi}}
(\frac{5p^4_{b,c}}{\alpha^2_{bc}}-26\frac{{p^2_{b,c}}}
{{\alpha_{b,c}}}-41)\nonumber\\
B_{b,c}&=&\omega_{b,c}(\frac{p^4_{b,c}}{4\alpha^2_{b,c}}-\frac{5p^2_{b,c}}{2\alpha_{b,c}}+\frac{15}{4})+(E_{p_{b,c}}-m_{b,c})\frac{\alpha_{b,c}}{2p^2_{b,c}}(\frac{p^4_{b,c}}{\alpha^2_{b,c}}-\frac{2p^2_{b,c}}{\alpha_{b,c}}+7)
\end{eqnarray}
For both 2s and 3s states: $$\phi_{b,c}=\tan^{-1}\frac{B_{b,c}}{A_{b,c}}$$ with respective $A_{b,c}$ and $B_{b,c}$

The binding energies of the constituent quark and antiquark for ground and orbitally excited $B_c$ and $B_c^*$ states can also be obtained by solving respective cubic equations with n=1,2,3 and l=0 representing appropriate bound-state conditions by putting the quantum number n=1,2,3 and l=0.

\begin{acknowledgments}
One of the authors Sonali Patnaik acknowledges the financial support and facilities provided by the authorities of Siksha O Anusandhan University, Bhubanaeswar, India to carry out the present work.
\end{acknowledgments}

\end{document}